\documentclass[preprint,showpacs,preprintnumbers,amsmath,amssymb]{revtex4}

\usepackage{graphicx}
\usepackage{dcolumn}
\usepackage{bm}

\begin{document}

\preprint{}

\title{ Constant rate shearing on two dimensional cohesive disks}
\author{N. Olivi-Tran$^1$,  O. Pozo$^2$, N. Fraysse$^2$}
\affiliation{$^1$ SPCTS, UMR-CNRS 6638, Ecole Nationale Superieure de
        Ceramiques Industrielles,47 av. A.Thomas, 87065 Limoges cedex, France\\
 $^2$  LPMC, UMR-CNRS 6622, Parc Valrose,Universite de Nice Sophia Antipolis,  06108 Nice cedex 2, France
}

\date{\today}

\begin{abstract}
We performed two-dimensional  Molecular Dynamics simulations of cohesive disks under shear.
The cohesion between the disks is added by the action
of springs between very next neighbouring disks, modelling capillary forces.
The geometry of the cell allows disk-disk shearing and not disk-cell wall
shearing as it is commonly found in literature.
Does a stick-slip phenomenon happen though the upper cover
moves at a constant velocity, i.e. with an infinite shearing force?
We measured the forces acted by the disks on the upper cover for different
shearing rates,
as well as the  disk  velocities  as a function of the distance to the
bottom of the cell.
It appears that the forces measured versus time present a periodic
behavior,very close to a stick slip phenomenon, for shearing rates larger than a given threshold. The disks' collective displacements in the shearing
cell (back and ahead) is the counterpart of the constant velocity of the upper cover
leading to a periodic behavior of the shear stress.
\end{abstract}


\maketitle
\frenchspacing   


\section{INTRODUCTION}
It frequently happens in nature that a system under continuous driving force
responds in an intermittent way. Time intervals when the system
is at rest and potential energy is accumulated alternate with active periods,
when the system relaxes and potential energy is decreased.
Since the simplest interaction leading to such behaviour is friction
between two moving objects, these phenomena are called stick slip
processes.
Since one of the dominating interactions in granular materials is friction
among the grains, it is not surprising that dense granular materials exhibit
various stick slip phenomena.

Cohesion is generally put into a granular system by the mean
of a liquid added to the grains. The presence of cohesion
adds a new dimension to the underlying physics of granular
materials. Experimentally and numerically, the physics of humid granular
media has only begun in the late 90's \cite{albert,barabasi,halsey,nedderman,horn,bocquet,fraysse,fraysse2,alonso,mason,nase,samadani,olivi1,olivi2,restagno} .

Here we will study numerically a two dimensional shearing cell 
containing mono-disperse cohesive disks.
The cohesion in our model is very weak: in comparison, this cohesion
may be experimentally obtained by the addition
of an undersaturated water vapor atmosphere
surrounding spherical beads. 
Hence, cohesion here is very different from that obtained
with addition of macroscopic quantities of liquid.
 In this cell, the shearing
rate is constant, i.e. the shearing driving force is infinite.
Thus, the system as a whole can not
be at rest.
We will analyze the behaviours of our two-dimensional cohesive disks,
and study the shear stress as a function of time, with a Molecular
Dynamics model \cite{olivi1,olivi2} of the experiment in order to answer the question:
does a stick slip phenomenon happen in cohesive disks dragged with an infinite shearing force?

\section{NUMERICAL MODEL}
The model we used here is a  version of Molecular
Dynamics for granular flow with cohesion in a two-dimensional
shearing cell \cite{olivi1,olivi2}. 
Particles are modeled as N disks that have equal density $d=2.2g.cm^{-2}$ and 
 diameters $d=0.2mm$. 

        The only external force acting on the system results from
gravity, $g = 981 cm.s^{-2}$.
        The particle-particle and particle-wall contacts are
described in the normal direction (i.e. in the particles'
center-center direction) by a Hooke-like force law. The normal force
is written:

\begin{eqnarray}
{\bf f}_n(i,j)=(-Yr_{eff}[\frac{1}{2} (d_i+d_j)-|{\bf_r}_{i,j}|]+\nonumber \\ \gamma \frac{m_{eff}({\bf v}_{i,j}.{\bf r} _{i,j})}{r_{eff}|{\bf r}_{i,j}|})\frac{{\bf r}_{i,j}}{| {\bf r}_{i,j}|}
\end{eqnarray}
where $Y$ is the Young modulus of the solid, $r_{eff}$ ($m_{eff}$) stands for the
effective radius (mass) of the particles $i$ and $j$, ${\bf v_{ij}}={\bf v_i}- {\bf v_j}$ is the relative
velocity of particle $j$ towards particle $i$.
$d_i$ (resp. $d_j$) is the diameter of particle $i$ (resp. $j$) and ${\bf r_{ij}}$
points from particle $i$ to particle $j$.
$\gamma$ is a phenomenological dissipation coefficient.

        We model the static friction force between particles by
putting a virtual spring at the point of first contact. Its
elongation is integrated over the entire collision time and set to
zero when the contact is lost. The maximum possible value of the
restoring force in the shear direction (i.e. in the plane
perpendicular to the normal direction), according to Coulomb's
criterion, is proportional to the normal force multiplied by the
friction coefficient $\mu$. It gives a friction force ${\bf f_s}(i,j)$ which is
written:
\begin{eqnarray}
{\bf f}_s (i,j)=-sign(f_f(i,j))min(f_f(i,j),\mu|f_n(i,j)|){\bf s}\\
{\bf f}_f(i,j):= - \int (\dot{r}_i- \dot{r}_j) {\bf s} dt
\end{eqnarray}
where $s$ stands for the unit vector in the shear direction.
        When a particle collides with the cylinder wall, the same
forces (1) and (2) act with infinite mass and radius for particle $j$.

 Capillary forces were modelled by adding a spring
force to the normal force when particles are in contact:
\begin{equation}
f_{cap}=Kr_{eff}\frac{1}{2} (d_i+d_j)
\end{equation}
where $K$ is the corresponding spring constant, which depends on the
surface tension and on the viscosity of the liquid.
When the distance between the surface of the particles is lower than 10 \% of the diameter
of the smallest particle, the value of the spring constant is multiplied by the distance
between the  particles. This additive
force is set to zero when the elongation of the virtual spring
reaches a maximum length of 10\% of the smallest particle diameter.

This cohesive force is a good model for capillary forces
between beads with nanoasperities: capillary bridges are
then located between two asperities belonging to two different
beads or between one asperity of one bead and one relatively flat surface of the other
bead. Therefore, the attractive force increases with an increasing
amount of liquid rendering our spring model for a cohesive force
relevant.

The disks are put in a shearing cell with blades, one can see an example
of this shearing cell in Figure 1. Periodic boundary conditions
are imposed on the left and right hand side of the cell.
Shear is applied on the disks by translating the upper cover at
constant velocity, whatever the resistance to translation 
of the granular medium.
The upper cover of the cell is allowed to undergo a vertical shift,
the magnitude of this shift depending on the weight of this upper cover, on the velocity of the upper cover and on the disks assembly dilation.

\section{RESULTS AND DISCUSSION}

 The parameters of our computations
were the followings: the velocity of the upper
cover was $v=0.2mm/s$ and the weight of the cover was equal to $=0.552g$, the mass of one disk
being equal to $2.76.10^{-3}g$. The length of the cell was $L=8mm$ and the height
of the blades
was $h=1mm$. We used $N=720$ disks.
The dissipation coefficient was equal to 0.7, the Young modulus was
$Y=2020 g.s^{-2}$ and the spring constant was $K=41.4 g.s^{-2}$.
Our model represents mesoscopic beads (i.e. larger
than particles in a powder) with a very weak cohesive
force, hence it is different from models with strong cohesion.
Hence, the quantity of disks used here is sufficient to analyze
their behavior on a mesoscopic level.

We computed the total force acted on the upper cover by the disk
assembly as a resistance to shear. For this, we added only the
horizontal coordinates of the forces acting on the cover.
This total force is similar to a stress.
In Figure 2, one can see an example of the evolution of this total
force as a function of time.
 The signal has been saved only when the permanent regime
was obtained, when the height of the cover and the mean signal
were steady.
 This signal appears to be very noisy: experiments with
a shearing cell with blades lead to similar irregular signals
(see \cite{pozo} ).
Anyway, it is very similar to a shear stress signal in a stick slip
phenomenon. We performed a direct Fourier transform on our signal, no characteristic peak appears
for this shearing rate which is the lowest that we tested.
Hence, the stress signal is not periodic. The stress signal becomes periodic
for shearing rates larger than $8mm/s$ and up to $120mm/s$, from Fourier
analysis of the stress signals corresponding to the these shearing rates
in our simulations.

In order to understand the origin of the intermittent signal,
we made an analysis of the internal geometrical structure i.e.
the locations of the disks during shearing in the cell.

In Figure 3 one can see the density of disks as a function
of the distance (in disk diameters) to the bottom of the cell.
We computed this density for two shearing rates: $v=0.2mm/s$
and $v=120mm/s$.
For $v=0.2mm/s$, it appears that the disks have a two dimensional
crystalline structure: we observe periodic peaks in the density of disks.
In the intervals between the peaks, the density increases slowly
as a function of the distance to the bottom.
All this means that the disks are structured in layers. 
For $v=120mm/s$ this layered structure is also present but
with lower densities (for each layer).
For this high shearing rate, the layered structure disappears at 
11 disks diameters from the bottom, and from then on decreases
slightly. While for the lower velocity, the layered structure
is maintained up to 16 disks diameters from the bottom,
and decreases only at a distance to the top corresponding
to the presence of the blades.
These layered structures are a consequence of the dimension of space: two-dimensional.
Indeed, the disks try to lower their resistance to shear and hence
take place in a layered structure where the layers are parallel
to the shearing force.

Let us compare the structure of the disks and the velocities of disks
inside the cell.
For that we took a snapshot of the instantaneous horizontal velocities of the disks
at a given time $t$ of the permanent regime.
Figure 4 is this snapshot for a shearing rate $v=0.2mm/s$
and Figure 5 is the snapshot for a shearing rate $v=120mm/s$.

Let us examine Figure 4 in the light of Figure 3.
Both figures have been set up with a shearing rate $v=0.2mm$.
In Figure 4, one can see that there remains the layered structure
of Figure 3: close to the bottom of the cell (close to 0 in disks diameters)
the velocities of the disks are periodically high, for distances up to 5 disk diameters
from the bottom. There does not appear clearly a shearing band:
the velocities of the disks increases slowly and regularly
from abscissa 5 to abscissa 25.
What is interesting to see is that the disks undergo negative and positive
velocities, that means that the disks follow the shear direction and also
the inverse direction.
This happens near the bottom of the cell but also in the region of the blades.
This is a consequence of the cohesion: when two disks collide, due to the spring
linking them, these two disks undergo a periodic oscillation backward
and forward. As all neighbouring disks are related by springs, there 
are collective oscillations of the disks inside one layer.
These collective horizontal oscillations disappear when  the layered structure
disappears.

Now, if we look at Figure 4 and Figure 3.
There are no more collective horizontal oscillations because of the high
value of the shearing rate ($v=120mm/s$).
It appears clearly a shearing band were the velocities of the disks increase
regularly form almost zero to the value of the shearing rate.
This shearing band has a width of 10 disk diameters.

The question remaining is: is there here a stick slip phenomenon
with a constant shearing rate i.e. a infinite shearing force?
In the light of the results that we obtained we can say
that the disks try to decrease their resistance to shear by
arranging themselves in layered structures parallel
to the shearing direction. Furthermore, due to the cohesion
given by springs between very next neighbours, the disks
undergo small oscillations, in the direction of the shear and
in its inverse.

The disks interact by the way of springs.
So, we may say that we obtained a stick slip
phenomenon in an assembly of disks put in a two-dimensional
shearing cell.
The stress signal that we obtained is the result
of a sticking stage, when the springs linking our disks
enlarge and the potential energy increases.
The slip stages correspond to the periods where the springs relax
and the potential energy is released. 

This is a mean behavior,
 and one cannot say whether
all springs relax at the same times.
From our Fourier analysis of the stress signal, we can say that,
for $v<8mm/s$ (no characteristic peak in the Fourier
curve) the relaxing and enlarging of the springs are not
correlated: the stress signal is non-periodic.
While for $8mm/s<v<120mm/s$, a characteristic peak appears
in the Fourier signal, the relaxing and enlarging of the springs
are correlated in time, and the stress signal is periodic.

We computed from this Fourier transform the characteristic
frequencies of the stick slip signal for $8mm/s<v<120mm/s$
and for two dimensions of the simulation box, as a function
of the shearing rate.
The first dimension corresponds to the preceding computations,
i.e. width $8mm$ and height of the blades $1mm$ with 720 disks. The second dimensions
are: width $12mm$ and height of the blades $1.5mm$ with 1080 disks.
Results are shown in fig.6.
We see in that figure that the stick slip signal frequency
depends on the dimensions of the simulation box.
As we used springs to model cohesion in the bulk
of our particles, we can say that the characteristic
frequencies of the oscillations of the assembly of springs
depend hence on the dimensions of the simulation box,
like the frequency of a unique oscillator in a box
of variable dimensions, even if we used periodic
boundary conditions. 

The characteristic frequencies depend also linearly
on the shearing rate.  This is the signature that the dissipation
of energy as a function of time is small compared to
the shearing rate: the amount of energy brought by the constant shearing rate
counterbalances the loss of energy due to dissipation when
two particles overlap.
Indeed, during each stick event, energy is accumulated by the way of the elasticity
of the disks and inside the springs (potential energy); as dissipation is
counterbalanced by the energy brought by the constant shearing rate,
the assembly of springs behaves as a unique spring with no dissipation.

\section{CONCLUSION}

We computed the behavior of two-dimensional  disks
with weak cohesion
in a shearing cell with blades.
Though the shearing rate is constant, we observe a typical stick slip stress
 signal. As the cohesion is added by the way of springs linking
very next neighbouring disks, we can say that the stick stage
corresponds to a collective enlargement of all springs where 
potential energy is accumulated.
The slip stage results from the relaxing of the springs.
Our stress signal becomes periodic for shearing rates values
larger than a given threshold.
The characteristic frequencies of the stick slip signal 
depend on the dimensions of the simulation box.
As the cohesion between the particles has been modelled
by springs, we can say that the whole assembly
of interacting particles behave as a global unique spring.
Hence each characteristic frequency depend on the shearing rate.
Moreover, as the shearing rate is constant , the amount of energy
necessary to counterbalance dissipation is always adequate,
leading to a linear evolution of the frequencies with shearing rate.

\begin{figure}
\includegraphics{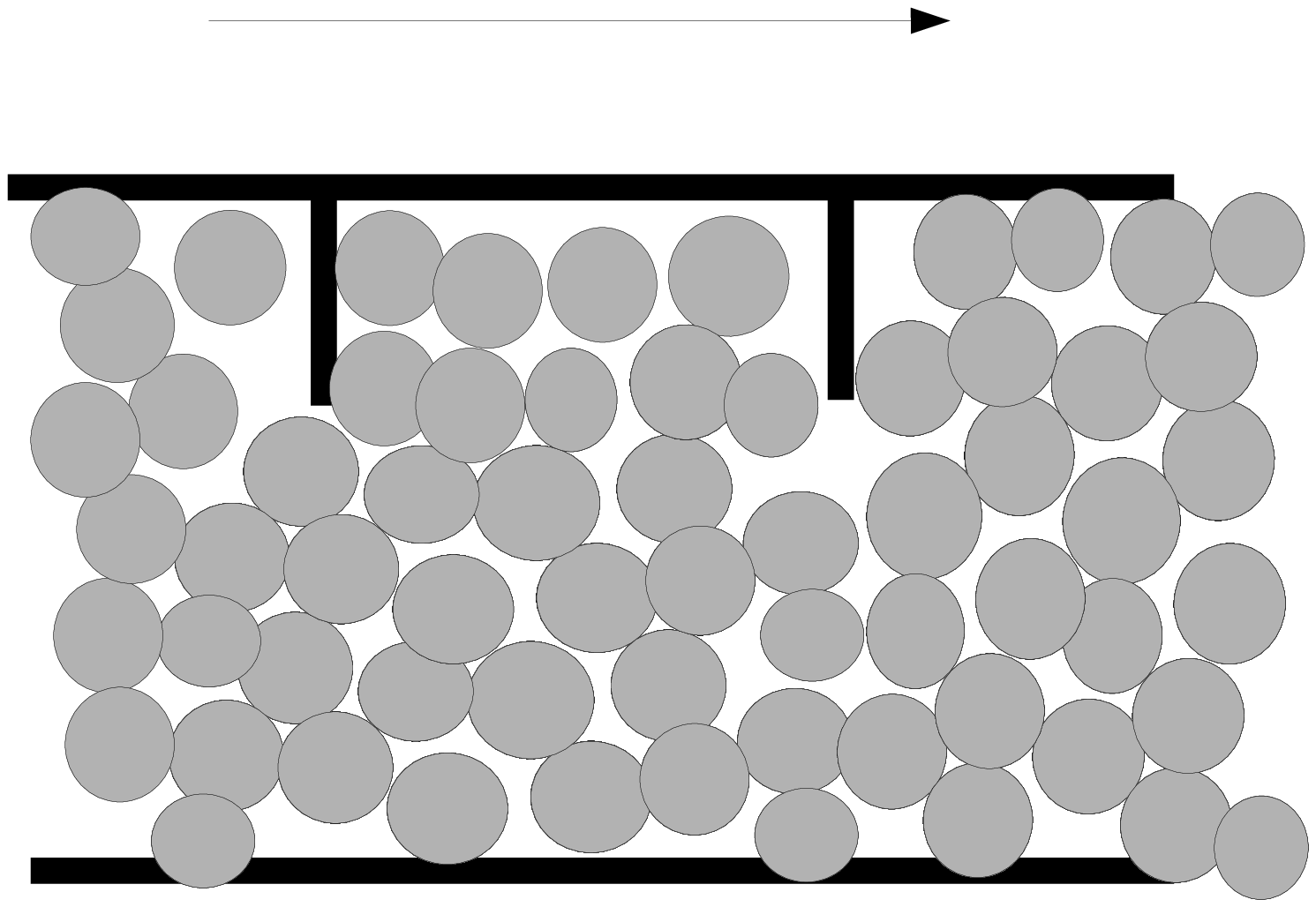}
\caption{Shearing cell; the arrow indicates the direction of shear
}
\end{figure}
\begin{figure}
\includegraphics{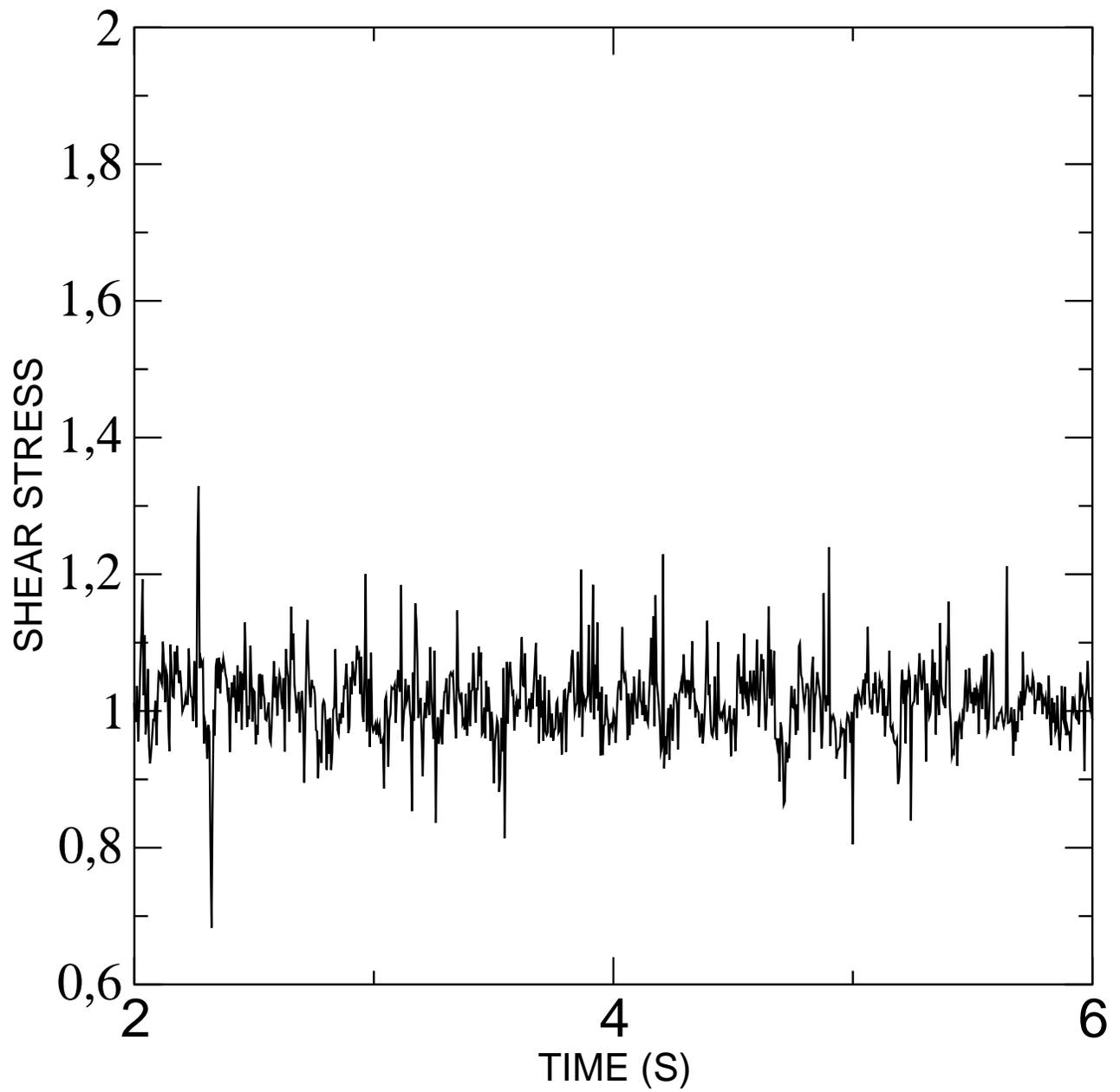}
\caption{Stress signal (arbitrary units) as a function of time (in $s$
for a shearing rate $v=0.2mm/s$
}
\end{figure}
\begin{figure}
\includegraphics{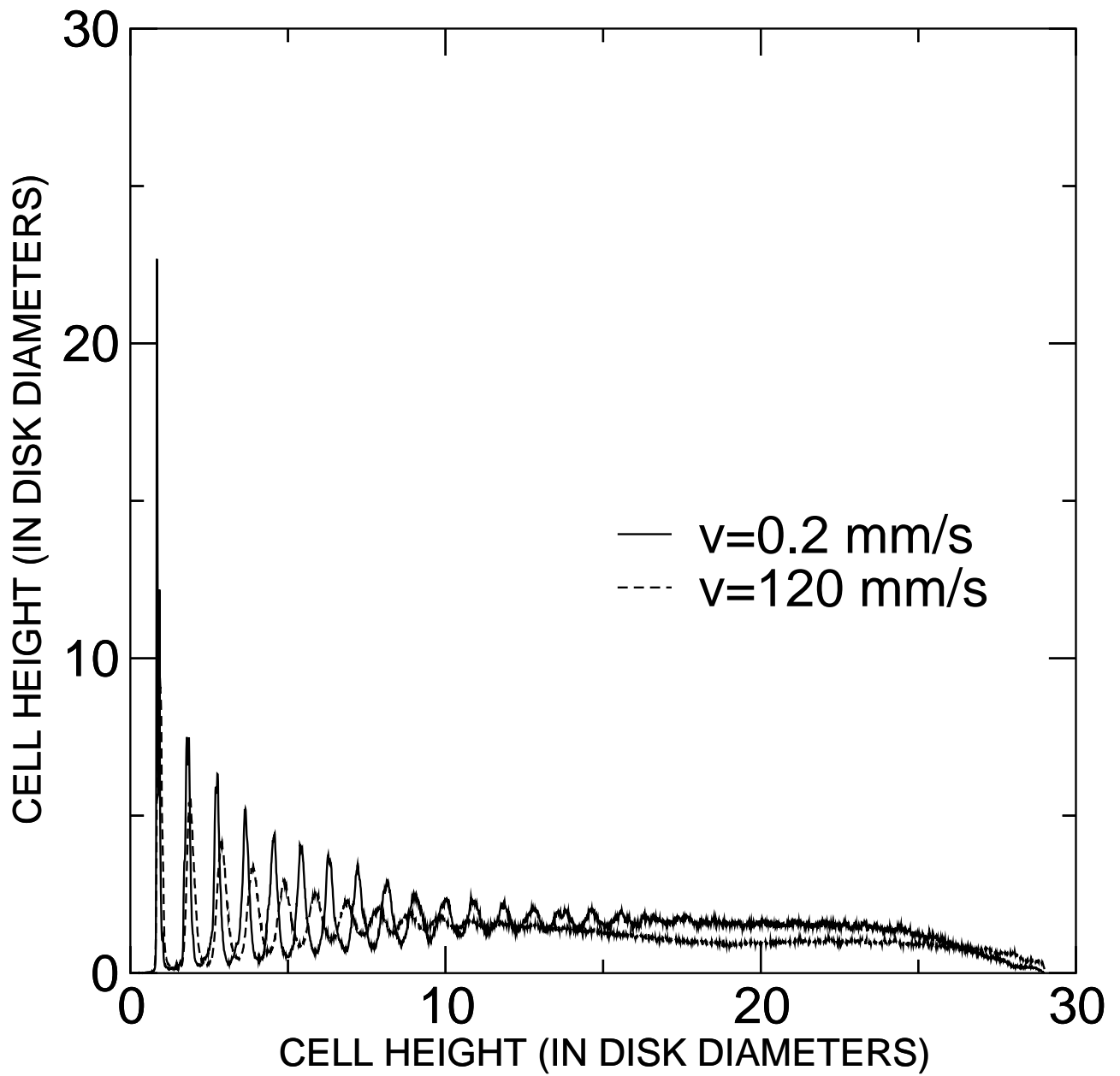}
\caption{Horizontal disk densities as a function of the distance
(in disk diameters units)
to the bottom of the shearing cell , for two shearing
velocities $v=0.2mm/s$ (continuous line) and $v=120mm/s$
(long dashed line)
}

\end{figure}
\begin{figure}
\includegraphics{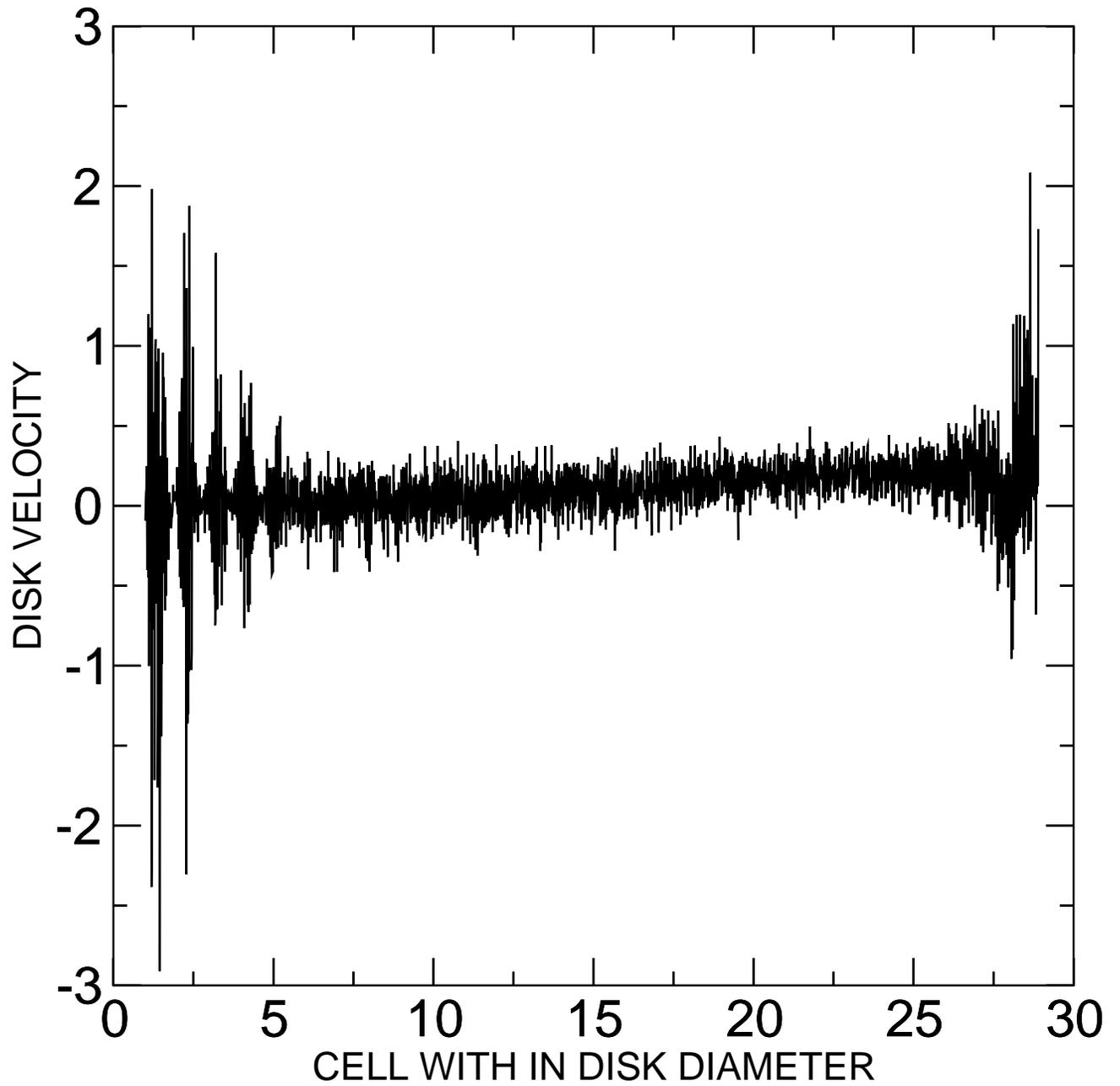}
\caption{Horizontal velocities (in $mm/s$) of the disks as a function of the distance
to the bottom of the cell (in disks diameters units) for a shearing rate
$v=0.2mm/s$
}
\end{figure}
\begin{figure}
\includegraphics{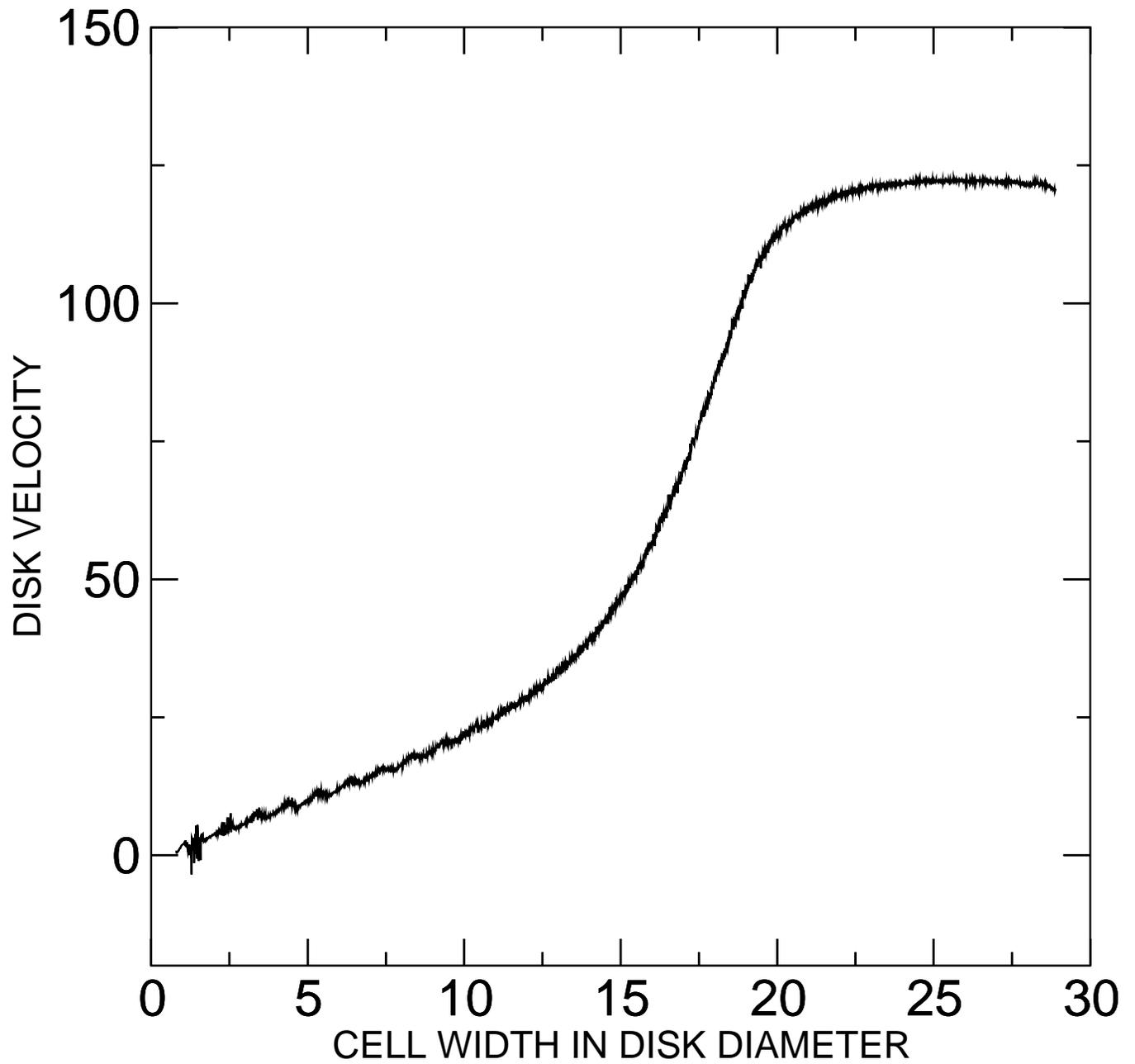}
\caption{Velocities (in $mm/s$) of the disks as a function of the distance
to the bottom of the cell (in disks diameters units) for a shearing rate
$v=120mm/s$
}

\end{figure}
\begin{figure}
\includegraphics{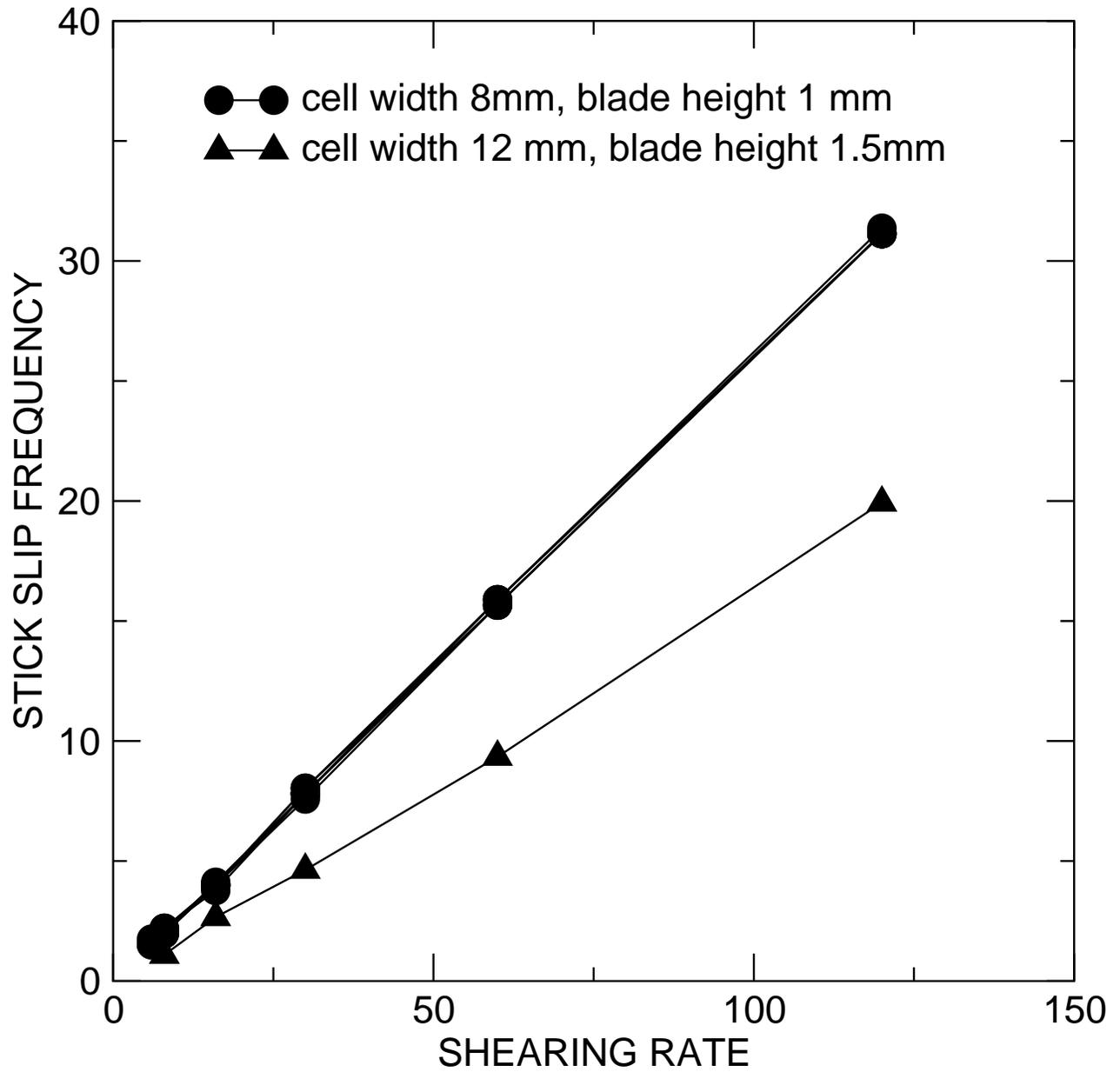}
\caption{Characteristic frequencies of the stick slip signal for two simulation
box dimensions (width=$8mm$ and height of blades=$1mm$ with 720 disks; width=$12mm$ and
height of blades=$1.5mm$ with 1080 disks), as a function of shearing rate.
}

\end{figure}


\begin{thebibliography}{99}
\bibitem{albert}{ Albert R, Albert I, Hornbaker D, Schiffer P and Barabasi AL{\it  Phys. Rev. E} {\bf 56} (1997) R6271}
\bibitem{alonso}{Alonso JJ, Hovi JP and Herrmann HJ {\it Phys. Rev. E} {\bf 58} (1998) 672}

\bibitem{barabasi}{ Barabasi AL,Albert R and Schiffer P  {\it Physica A} {\bf 266} (1999) 366}
\bibitem{bocquet}{Bocquet L, Charlaix E, Ciliverto S and Crassous J  {\it Nature} {\bf 396} (1998) 735}
\bibitem{fraysse}{Fraysse N, Thome H and Petit L , In Behringer RP and Jenkins JT (Eds.)                                                                     {\it Powder and Grains 1997} Rotterdam; Balkema AA}
\bibitem{fraysse2}{Fraysse N, Thome H and Petit L {\it Eur. Phys. J. B} {\bf 11} (1999) 615}

\bibitem{halsey}{ Halsey TC and Levine AJ  {\it Phys. Rev. Lett. } {\bf 80}(1998) 3141}
\bibitem{horn}{Hornbaker DJ, Albert R, Albert I, Barabasi AL and Schiffer P  {\it Nature} {\bf 387} (1997) 765}
\bibitem{mason}{Mason TG, Levine AJ, Ertas D and Halsey TC  {\it Phys. Rev. E } {\bf 60} (1999) 50
44}

\bibitem{nase}{Nase ST,Vargas WL, Abatan AA, McCarthy JJ  {\it Powder Techn. } {\bf 116} (2001) 214}

\bibitem{nedderman}{ Nedderman RM ,Statics and kinematics of granular materials, (Cambridge,1992, Cambridge Univ. Press)}
\bibitem{olivi1}{ Olivi-Tran N, Fraysse N, Girard P, Ramonda M and Chatain D {\it Eur. Phys. J.
B} {\bf 25} (2002) 217}
\bibitem{olivi2}{Olivi-Tran N, Pozo O and Fraysse N  {\it Model. and Simul
. in Mat. Sci. and Eng.}{\bf 12} (2004) 671}
\bibitem{restagno}{Restagno F,Bocquet L and Charlaix E  {\it Eur. Phys. J. E }{
\bf 14} (2004) 177}
\bibitem{samadani}{Samadani A and Kudrolli A,  {\it Phys. Rev. Lett. } {\bf 85} (2000) 5102}
\bibitem{pozo}{Pozo O, Fraysse N and Olivi-Tran N  {\it Powder and Grains 2005} to appear}

\bibliography{mybib} 
\end{thebibliography}
\end{document}